# FAMILY OF 2-SIMPLEX COGNITIVE TOOLS AND THEIR APPLICATIONS FOR DECISION-MAKING AND ITS JUSTIFICATIONS

Yankovskaya Anna[1] and Yamshanov Artem[2]

[1]Tomsk State University of Architecture and Building, Tomsk, Russia
Tomsk State University of Control Systems and Radioelectronics, Tomsk, Russia
National Research Tomsk State University, Tomsk, Russia
Siberian State Medical University, Tomsk, Russia
`ayyankov@gmail.com`
[2]Tomsk State University of Control Systems and Radioelectronics, Tomsk, Russia
`yav@keva.tusur.ru`

## ABSTRACT

*Urgency of application and development of cognitive graphic tools for usage in intelligent systems of data analysis, decision making and its justifications is given. Cognitive graphic tool "2-simplex prism" and examples of its usage are presented. Specificity of program realization of cognitive graphics tools invariant to problem areas is described. Most significant results are given and discussed. Future investigations are connected with usage of new approach to rendering, cross-platform realization, cognitive features improving and expanding of n-simplex family.*

## 1. INTRODUCTION

One of the most important and difficult problem in creating the intelligent system for data and knowledge analyzing, and decision-making is the development of a software library for representation of the results in a user friendly and clear view for a wide group of users. It is particularly important when technologies are rapidly changing from data-limited area to data-driven analysis-limited area [1]. The ability to generate big databases of experiments is far ahead from the ability to analyze and visualize these databases. One solution of these problems is usage of the cognitive graphic tools.

An important contribution to the development of the cognitive science was made by R. Axelrod [2], R.G. Basaker [3], D.A. Pospelov [4-6], A.A. Zenkin [7-8], V.F. Khoroshevskiy [9], B.A. Kobrinskiy [10], A.E. Yankovskaya [11-13]. Namely they, who saw the potential of using cognitive tools in the various problem areas, and despite the meager technical capabilities which were available at that time, first steps in the development of the cognitive graphics tools were made, and thereby a new research direction was formed [7, 10]. These tools are very effective for the interpretation of analyzed data and knowledge, decision-making and its justifications for users who are specialized in different problem areas but are not specialists in algorithms of data analysis and knowledge inference which are used in intelligent systems. The cognitive graphic tools are important link between a big amount of data and understanding these data. The application of the cognitive graphic tools makes possible to understand a lot of processes occurring at the lowest level which was not understandable before and revealing of different new regularities and connections between different factors and events in a wide variety of problems and cross-disciplinary areas. The cognitive graphic tools are used in different intelligent systems for information data and knowledge structures analyzing, for revealing regularities of different kinds and decision-making and its justification. They also can be used in the intelligent learning-testing systems for teaching and learning activities optimization, for visualization and forecasting of learning process results and etc. But the development of these tools for each problem area is very time-consuming and expensive. Thus, the cognitive tools which are invariant to different problem areas were developed [14-15]. The specificity of these tools application is the software realization does not aim at a concrete problem area and is realized as visualizing plugins for intelligent instrumental software (IIS) IMSLOG [16] which is used for constructing specific applied intelligent systems. These visualization plugins include the cognitive graphics tools for visualizing information structures, different kinds of regularities, decision-making and its justification etc.

Usage of these tools is not only actual for the parameters analysis of the different non-changing states of objects for decisions justification. It is also actual for analysis and visualization of the dynamic processes. Despite the fact that visualization is not necessary for decision-making, it simplifies the analysis of information and provides possibility for the best decision-making. For example, an intelligent system user in the area of a concrete discipline teaching (a lecturer) considers different aspects of respondents teaching during a time of a real exam or an intelligent system user in the area of the diseases diagnosis (a doctor) considers the previous stages of patients examinations.

The given article continues the research of the cognitive tools application based on the n-simplex [15]. Undoubtful advantages of the cognitive tools based on n-simplex are invariability to the problem areas. It should be pointed out that using the modern technologies for creation of cognitive graphic tools allows to get different cognitive tools: such as desktop application (application for desktop PC), applications for smartphones and pads, WEB application.

Mathematic basis of an object under study representation in n-simplex is described and basis of representation of a process under study in 2-simplex prism is given. Examples of 2-simplex prism application in developed and developing intelligent systems are presented. Further study directions are proposed.

## 2. FAMILY OF 2-SIMPLEX COGNITIVE TOOLS

### 2.1. Cognitive Tool 2-Simplex

Firstly, transformation of features space in patterns space based on the logical-combinatorial methods and properties of n-simplex are suggested in the publication [17]. System of visualization TRIANG for decision-making and its justifications with cognitive graphics [18] is constructed on the base of 2-simplex with usage of the following theorem [17].

*Theorem.* Suppose $a_1, a_2, ..., a_{n+1}$ is a set of simultaneously non-zero numbers where $n$ is the dimension of a regular simplex. Then, there is one and only one point that following condition $h_1 : h_2 : ... : h_{n+1} = a_1 : a_2 : ... : a_{n+1}$ is correct, where $h_i (i \in 1, 2, ..., n+1)$ is the distance from this point to $i$-th side [17-18].

Coefficient $h_i (i \in 1, 2, ..., n+1)$ represents the degree of conditional proximity of the object under study to $i$-th pattern [17]. The advantage of this fact is the n-simplex possesses the property of the constancy of the sum of distances ($h$) from any point to each side and the property of ratios preservation $h_1 : h_2 : ... : h_{n+1} = a_1 : a_2 : ... : a_{n+1}$. Distances $h_i$ are calculated on the basis of coefficients $a_i (i \in 1, 2, ..., n+1)$ and normalization operations from following relations

$$\begin{cases} H = \sum_{i=1}^{n} h_i \\ H = A \sum_{i=1}^{n} a_i \\ \dfrac{h_1}{a_1} = \dfrac{h_2}{a_2} = ... = \dfrac{h_n}{a_n} \end{cases} ,$$

where $A$ – scaling coefficient) by the formula
$$h_i = A \cdot a_i , \ i \in \{1, 2, ..., n\}$$

This theorem was used in more than 30 applied intelligent systems and in three instrumental tools of revealing different kind of regularities and making of diagnostic, classifications organization and control decisions and their justification.

The main function of n-simplex is a representation of a disposition of object under study among other objects of a learning sample. Additionally, n-simplex has other useful functions for a decision-making person. One of these functions is a representation of some numerical values, for example, an admissible error of recognition preassigned by the user. Example of 2-simplex is shown on Figure 1.

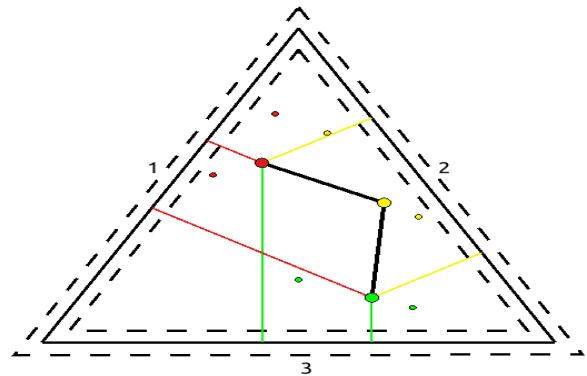

Figure 1. Example of 2-simplex

The sides of 2-simplex (triangle edges) are associated with patterns (classes), circles with big radius are the objects under study and circles with small radius are learning sample objects. The distance from the object to a side is directly proportional to the proximity of the object to the pattern corresponding to the side. The distance for object under study is displayed as color perpendicular lines to 2-simplex sides (red, yellow, green). The color of the object under study or objects from learning sample is mapped to a pattern which is revealed for a specific object. Line segments between objects represent the dynamics of process under study, for example, they can represent changing of a student knowledge level. An object color is mapped with an associated pattern (the nearest pattern or pattern determined by an expert). Digits are mapped with a pattern and are placed on associated sides. It is not a usual working mode because it makes the image more complex, and it is not necessary because the association between the side and the pattern can be determined by a color of the perpendicular line from the point to the side. So, for usual working mode it is suggested to hide these digits. But for the first look or demonstration it can be quite useful.

## 2.2. Cognitive Tool 2-Simplex Prism

The cognitive tool "2-simplex prism" (Figure 2) is based on 2-simplex and represents the triangular regular prism which contains in basics and cuttings 2-simplexes which are corresponded fixed time moments.

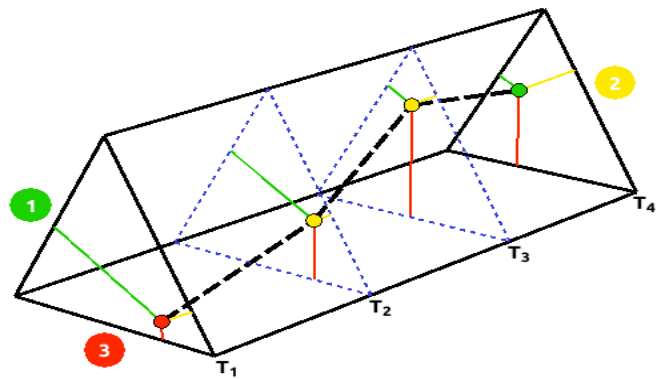

Figure 2. Example of 2-simplex prism

Distance from the base of the prism to $i$-th 2-simplex $h_i'$ corresponds to the fixation moment of object under study features and it is calculated based on the following formula:

$$h_i' = H' \cdot \frac{T_i - T_{min}}{T_{max} - T_{min}},$$

where    $H'$ – length of 2-simplex prism preassigned by a user and corresponded to the study duration,

$T_i$ – timestamp of features fixation of object under study for $i$-th examination,

$T_{min}$ – timestamp of features fixation of object under study for the 1-st examination,

$T_{max}$ – timestamp of features fixation of object under study for the last examination.

Because 2-simplex prism is based on the 2-simplex description of all 2-simplex objects is also correct for 2-simplex prism.

2-simplex prism allows representing visually as well the dynamic processes as the modeling one or another process, which is necessary for a big amount of problems and cross-disciplinary areas: medicine, economy, genetics, building, radioelectronics, sociology, education, psychology, geology, design, ecology, geo-ecology, eco-bio-medicine etc.

## 3. SOFTWARE IMPLEMENTATION

Software prototypes of the described cognitive tools are implemented with using C# language. Source information for visualization (named listing of n-simplex or LNS) is a JavaScript code which describes n-simplex: objects under study, objects of learning sample objects, links between points and other parameters that are necessary for tuning a size of n-simplex, point of view, type of transformation etc. All described cognitive graphic tools implemented as two libraries: first for 2D visualization (2-simplex and 3-simplex unfoldings) and second for 3D visualization (3-simplex, 2-simplex prism). They have different features, capabilities and visualization parameters but can use one universal LNS. Both n-simplex library parse and visualize any LNS which describe only objects and which do not describe visualization parameters. If it is necessary to set additional visualization parameters for any library and keep it compatible with another library must be used special language constructions. The libraries visualize n-simplex only from LNS located in memory. Earlier there was a function to visualize n-simplex from LNS located in file, but it was excluded because such functionality is not supported on all platforms (for example, java-script for web client) and this functionality not necessary for most intelligent systems and quite trivial in realization. Library "Jint" was used for a LNS parsing. The second version of LNS language was implemented, but the new language is also simple for a code generation as before. Additionally new version of LNS language has more powerful and clearer syntax for creating and editing n-simplexes by user. The output of the library is a bitmap image visualized by GDI+ library. The fragment of the universal LNS language for a simple 3-simplex is given below.

```
try { setView(0); } catch (ex) { }
try { setTransform(2); } catch (ex) { }
try { setViewPort(15, 80); } catch (ex) { }

var size = 200,
```

```
    delta = 40,
    colors = ["#E01B1B", "#E0841B", "#F7F307", "#07F70B"],
    color = "#000",
    dashPattern = [1];

addTetraedron(color, 3, dashPattern, size);

addIJK(color, 2, dashPattern, size, [1, 2, 4, 8], colors);
addIJK(color, 2, dashPattern, size, [8, 4, 2, 1], colors);

addPoint(colors[0], 6, "Circle", size, [1, 2, 4, 8]);
addPoint(colors[1], 6, "Circle", size, [8, 4, 2, 1]);

addPath(color, 4, dashPattern, size, [
    [1, 2, 4, 8],
    [4, 4, 4, 4],
    [8, 4, 2, 1]
]);
```

The visualization and computation algorithms used in both libraries are identical and they are described in the paper [19]. There were various attempts to improve the visualization libraries during this year. The most significant results are the following:

1.  An attempt to change a raster renderer to a vector renderer proposed in [18] has failed, because a) partitioning of the objects into smaller entities and sorting them were too time-consuming tasks for rendering in real time to implement the interactive capabilities; b) a developing a vector renderer needs a lot of time and it is too difficult for a quick prototyping.

2.  Descriptive language for n-simplex was changed from our custom language to java-script. It was the first step to make library cross-platform and allows integrating our cognitive graphic tools in web applications.

3.  Big part of both libraries was universalized and moved to another library. The library contains the code for rendering graphics primitives, algorithms for color transformation, parsing modules etc. This step makes support and development of the libraries simpler and will simplify the porting of the libraries to java-script language which is necessary for an embedding into web-applications.

4.  Results of experiments with shaders, OpenGL ES and WebGL were obtained. They show that implementation of all desired functions of planned vector render is also possible with a raster renderer with usage of shaders. Moreover, a shader program is performed using hardware acceleration and all planned tasks can be implemented without losing rich interactive functionality. The implementation of all planned ideas will provide: translucent faces, intelligent layout of signatures, identification of objects at the specified point etc.

The most important advantages of a raster render are rich of graphic abilities (color, layout, complexity etc.) and rich of interactive functions. The most significant disadvantage of a raster render is the impossibility of scaling, but it is not important because n-simplex can be rendered directly in a desired size.

# 4. EXAMPLES OF APPLICATIONS IN DIFFERENT PROBLEM AREAS

A matrix approach to representation of data and knowledge which contains description matrix of objects in space of characteristic features and distinction matrix represented partitioning of objects on equivalence classes for every mechanism of classification are used in intelligent systems (IS) developed by us [14, 20]. The set of all non-regular matrix rows distinction matrix corresponds to set of revealed patterns.

A pattern is a subset of description matrix rows with equal of characteristic features values. In IS based on a logical-combinatory (l-c), a logical-probabilistic (l-p) and a logical-combinatory-probabilistic (l-c-p) methods of test pattern recognition and decision-making with usage of cognitive tools [11, 14, 20] mathematical basis for calculation of conditional proximity coefficients of the object under study to $i$-th pattern is proposed. For l-c methods these coefficients are corresponded to a ratio between proximity coefficient object under study x to a pattern k and proximity coefficient object into a pattern k, for l-p and l-c-p these coefficients corresponded with probability to make decision for studied patterns Software implementation of these models for IS include development of corresponded mathematical apparatus for transformation of features space to patterns space which is described in section 2.

## 4.1. Application in Educational Area

This chapter describes visualization of testing knowledge result in learning-testing system with estimation coefficients usage [21]. In learning-testing system developed by us respondent after studying selected discipline, should pass mixed diagnostic test. During solution of this test respondent actions map (RAM) is forming. After completed all the questions, he respondent RAM projected in the set of predetermined valuation coefficients that determine how well the respondent cope with different tasks based on the following abilities (skills):

1.  storage and reproduction of the material in unmodified form;

2.  the reproduction of the material in modified form;

3.  extraction of new knowledge based on the studied material;

4.  problem solving, etc.

For example, for development of client-server software system with multimedia capabilities set of evaluated factors reasonable transformations to the following parameters:

1.  the solution of problems requiring high concentration;

2.  decision-making of non-trivial tasks;

3.  fast learning and knowledge of a large number of technologies.

It should be noticed that the different set of these parameters ($a_1, a_2, a_3$) can be transformed in same distances $h_1, h_2, h_3$ in case when sums of $a_i$ for different sets are equal. So for that and similar cases it is necessary to introduce the new parameter: a color saturation of point corresponded to the sum of $a_i$: $a_1 + a_2 + a_3$.

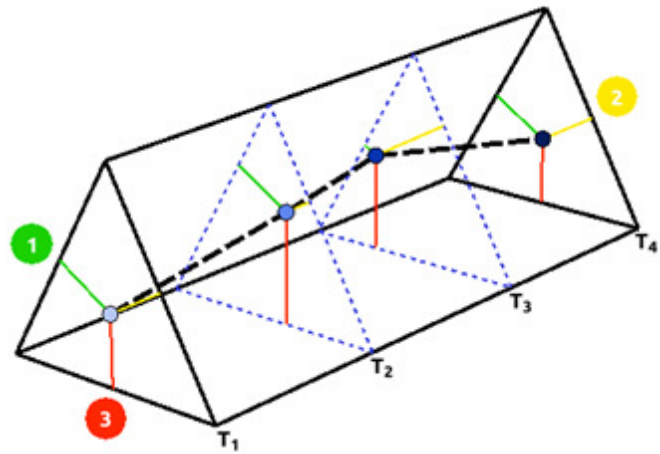

Figure 3. Example of 2-simplex prism usage for learning-testing systems

2-simplex prism allows represent dynamics for ability development a respondent or a group of respondents. But it should be noticed that representation of big group of respondents with usage 2-simplex prism can be too complex and inconvenient.

## 4.2. Decision Support for Diagnostics and Correction of Psychosomatic Disorders

Recently a number of studies were performed in the field of cognitive behavioral therapy and occupational stress [22–25] in order to identify the persons' psychological state, detect the essential patterns of the problem and correct persons' behavior, providing individual trajectory.

Versatile cognitive tools for identifying the attractiveness of organization are proposed in [26]. Such phenomenon as burnout is thoroughly discussed in [27] and the means of its prevention are considered. The platform for cognitive stimulation, maintenance and rehabilitation for health professionals is proposed in [28]. It allows to define and customize therapeutic intervention programs for cognitive rehabilitation or maintenance, relevant to multiple types of cognitive disorders especially for aged people. In [29] the fact that all processes in psychology happen in time and evolve over time is outlined. Therefore the versatile tool for processes estimation in dynamics is needed for wide range of applications, especially in psychology and medicine. This is especially relevant for person behavior prediction and timely correction in order to prevent their dissatisfaction and turnover intentions [30]. The urge of cognitive testing tools for aged people with disorders of cognition is discussed in [31].

In this paper we propose the cognitive tool designed for the intelligent system for diagnostic and intervention of an organization stress (DIOS) [32]. The following examples of 2-simplex prism usage in area of diagnostic and intervention of organization stress (OS) is given. The system under study is based on the idea of three-stage diagnostic. Our original questionnaire for intervention choosing for every stage (1 - alarm, 2 - resistance, 3 - exhaustion) [33] and fuzzy and threshold logics [34]. The idea of three-stage diagnostic allows in a short period of time to make a differential medical care for a patient with diagnosed OS.

DIOS use special questionnaire for the express-diagnostic of organization stress including a question for three stage OS: stage 1 is an alarm stage, the stage 2 is the resistance stage, stage 3 is a stage of exhaustion. This questionnaire is based on a Selye conception [35]. 7 features (symptoms) are used for revealing OS in stage 1 and stage 2, 8 features are used in stage 3. Allowed values for each feature are nothing - 0.0, seldom - 0.25, sometimes - 0.5, often - 0.75, constantly - 1.0. Analysis of features values allow to perform an express-diagnostic for revealing OS on every stage.

After the performing of a test system handle patient questionnaire and output diagnostic result of OS for every stage. These results are transferred to a module of visualization and justification. If study of OS recovering dynamics is not required then for justification of result cognitive tools 2-simplex is used. If the dynamics is required then for such purpose 2-simplex prism is used. The represented dynamics of OS diagnostic with a usage of 2-simplex prism allow examine accuracy of revealed diagnosis and selected medical intervention.

2-simplex prism can represent only relation for three patterns, but for some cases more patterns will be necessary. Example of this case is developed by us IS system DIOS which operate with 4 patterns: 3 stages of OS (alarm, resistance, exhaustion) and its absence. Current experiments show that most reasonable way to represent dynamics relating to such cases is usage of additional 2-simplex prism. For DIOS two 2-simplex prisms are used: the first represent object under study relate stages 3-1, the second relate stages 2-1 and OS absence. It should be noticed that both 2-simplex prism are not necessary for all cases. If dynamics of a patient recovering is limited by stages 3-1 or by stages 2-0 and absence it is reasonable to use only one 2-simplex prism.

The follow example of patient dynamics which cannot be limited by one 2-simplex prism is presented. 5 tests for revealing OS were performed. The first 2-simplex prism represents results for 1-4 tests, the second represents for 3-5 tests.

The first 2-simplex prism (Figure 4) represents transformation process from the stage of exhaustion (stage 3) to alarm stage (stage 1).

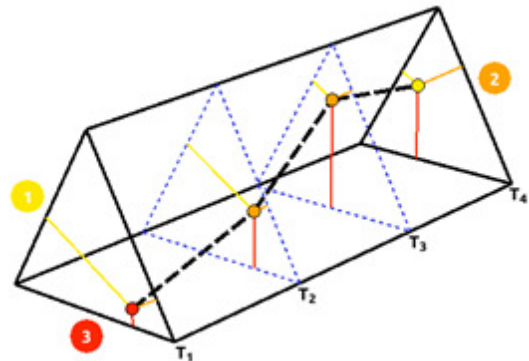

Figure 4. Visualization of tests 1-4 in 2-simplex prism

The first test ($T_1$) reveals a level between the stage of exhaustion and the resistance stage and prepotency of the stage of exhaustion over the resistance stage. The second test ($T_2$) reveals that illness is decreasing from the exhausted stage (pattern 3) to the resistance stage (pattern 2). The third test ($T_3$) reveals that illness is decreased to a level between the resistance stage (pattern 2) and alarm stage (pattern 1). The fourth test ($T_4$) reveals prepotency of the alarm stage (pattern 1).

The second 2-simplex prism (Figure 5) represents the transformation process from the resistance stage (pattern 2) to the absence of stress (pattern 0).

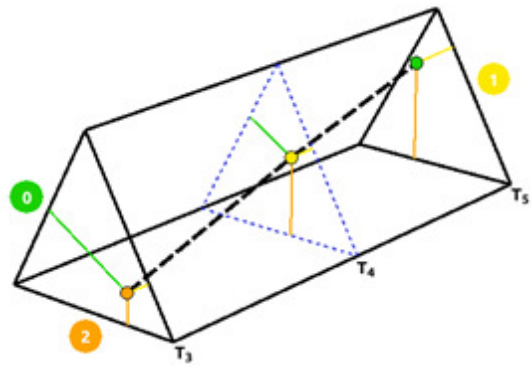

Figure 5. Visualization of tests 2-5 in 2-simplex prism

The fifth test ($T_5$) reveals the absence of the stress organization.

It should be noticed that the cognitive property of color are used in 2-simplex prism to represent dangerous of diagnoses and patterns.

## 4.3. Cognitive Modelling

The cognitive modelling of a decision-making in the artificial intelligent systems is the one of most important directions for creating intelligent systems (IS) in some priority areas of science researches and developing as medicine, psychology, sociology, environmental protection, energetics, systems of transport and telecommunication, control systems etc. Manipulation of some parameters of an object under study and using cognitive tools of decision-making and its justifications in IS we can perform a cognitive modelling base on the different kinds of knowledge representations.

For the cognitive modelling of decision-making only some part of the cognitive tools can be used: these tools should allow to visualize spatial relationship object under study and save sum of distances from any object to patterns (decisions made) and relations between them in process of transformation from space to feature space. Made normalization coefficient of the degree of conditional proximity (for l-c recognition) and set probability of decision making for studied objects (for l-p, l-c-p recognition), representation of object on geometrical figure allows to visualize patterns, see object position among other object from knowledge base and its proximity to a particular pattern. Nowadays in our IS for a cognitive modelling purpose 2-simplex is used, but using of a 2-simplex allows to make a modelling process more clear. The cognitive modeling of decision-making is performed by applying of some actions associated with, for example, using of therapeutic interventions in medicine, the adoption of measures to ensure environmental safety etc. An object answers on some of these actions (some values of some parameters of the object are changing) and their location on 2-simplex is changed. The result of a modeling gives us the dynamic image of an object location and relations between objects in a moment of actions and is represented with the cognitive tools by points which are sequentially connected by a polyline.

The example of prediction and therapeutic intervention with using of 2-simplex prism is presented in Figure 6. After the first patient examination ($T_1$) the diagnosis is revealed - stage 3 (exhausted) of organization stress and strategy of recovering is changed. With usage of mathematical model of a patient and recovery process it is possible to predict progress of patient recovery, which is shown in Figure 6 as polyline of a light-blue color.

After the second examination ($T_2$), a progress of a psychology stress recovering is diagnosed - organization stress is moved from stage 3 to stage 2 (resistance), but a real progress is worse than the predicted progress. At this moment the doctor has two different strategies for continuation of recovering: purple and blue. The doctor uses this cognitive representation of different recovering strategies. He can choose one which gives the best result in the future. At this moment a strategy associated with a polyline of a purple color is more reasonable, because this strategy is applied to a patient. The model predict full patient recovery until the moment of the next examination ($T_3$).

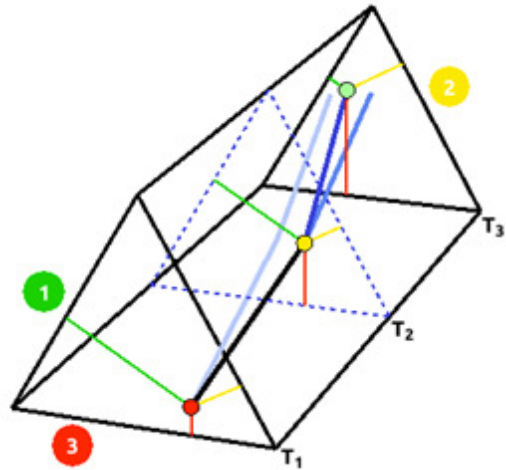

Figure 6. Visualizing of a modeling result in 2-simplex prism.

The cognitive modelling of a decision-making is based on mathematical and computer methods with applying tools form n-simplex family allows to optimize a choice of influence of object under study in accordance with a dynamical model of parameters changing.

Applications of 2-simplex is not limited by the above-mentioned examples.

## 5. CONCLUSIONS

The description of 2-simplex prism and examples of its applications are presented. The most important advantage for the information visualization in 2-simplex prism is the opportunity to analyze in dynamics the object under study over time. It allows the users to make decisions, to justify them and to analyse changes of parameters of object under study.

Application of the cognitive tools can be performed for any problem and cross-disciplinary areas in which it is necessary to make decisions about relation of object under study to one or another pattern (class) in fixed time moment or time interval and justify these decisions. Unlike any of the previously developed cognitive tools based on n-simplex [15, 36], 2-simplex prism allows to study objects dynamically on the time range interested for a user.

Development of cognitive graphics tools invariant to problem areas, their cross-platform realization and their integration in intelligent systems are presented. The implementation of some previously planned steps greatly reduced the time and labor costs for cognitive tools development and improved the human-to-machine interface.

In future we suppose to completely rewrite raster renderer using shaders technology for developing cross-platform realization which can be integrated in web-, desktop- and mobile-applications; to develop interactive features and cognitive properties of described cognitive graphic tools and to expand family of n-simplex.

## ACKNOWLEDGEMENTS

Supported by Russian Foundation for Basic Research, project 13-07-00373a, 13-07-98037-r_sibir_a, 14-07-00673 and partially by Russian Humanitarian Scientific Foundation project 13-06-00709.

## REFERENCES

[1]  S. R. Niezgoda, A.K. Kanjarla, S.R. Kalidindi Novel microstructure quantification framework for databasing, visualization, and analysis of microstructure data. Integrating Materials and Manufacturing Innovation 2013, 2:3 doi:10.1186/2193-9772-2-3.

[2]  R. Axelrod The Structure of Decision: Cognitive Maps of Political Elites. Princeton University Press, 1976.

[3]  R.G.Basaker, T.L.Saati Finite Graphs and Networks: An Introduction with Applications. Research Analysis Corp., Mc Graw Hill Company, NY-London-Toronto, 1965.

[4]  D.A. Pospelov Cognitive Graphics is a window into the new world. Software products and systems, 1992, 4-6 (in Russian).

[5]  D.A. Pospelov Ten "hot spots" in research on artificial intelligence. Intelligent systems (MSU), vol. 1(1-4), pp. 47–56, 1996 (in Russian).

[6]  D.A.Pospelov, L.V.Litvintseva How to combine left and right?. News of Artificial Intelligence, N2, 1996. (in Russian).

[7]  A.A. Zenkin Cognitive Computer Graphics. M.: Nauka, 1991 (in Russian).

[8]  A.A. Zenkin Knowledge-Generating Technologies of Cognitive Reality. News of Artificial Intelligence, N2, pp. 72-78, 1996 (in Russian).

[9]  V.A.Albu, V.F.Khoroshevskiy COGR – Cognitive Graphics System, Design, Development, Application. Russian Academy of Science Bulletin. Technical Cybernetics - 1990. - № 5 (in Russian).

[10] B.A. Kobrinskiy Why should we take in account imaginary thinking and intuition in medical expert systems. Artificial Intelligence – 96. Proceedings of the 5th National Conference with International Participation. Volume II. – Kazan, 1996 (in Russian).

[11] A.E. Yankovskaya Decision-making and decision-justification using cognitive graphics methods base on the experts of different qualification. Russian Academy of Science Bulletin, Theory and Control Systems, № 5, pp. 125-126, 1997 (in Russian).

[12] A. Yankovskaya, D. Galkin Cognitive Computer Based on n-m Multiterminal Networks for Pattern Recognition in Applied Intelligent Systems. Proceedings of Conference GraphiCon'2009. – Moscow.: Maks Press, 2009. – pp. 299-300.

[13] A. E. Yankovskaya, D. V. Galkin, G. E. Chernogoryuk Computer Visualization and Cognitive Graphics Tools for Applied Intelligent Systems. Proceedings of the IASTED International Conferences on Automation, Control and Information Technology, v.1. – 2010. – pp. 249-253.

[14] A.E. Yankovskaya Logical tests and means of cognitive graphics. Publishing house: LAP LAMBERT Academic Publishing, 2011 – 87 c. (in Russian).

[15] A.E. Yankovskaya, N.M. Krivdyuk Cognitive graphics tool based on 3-simplex for decision-making and substantiation of decisions in intelligent system. Proceedings of the IASTED International Conference Technology for Education and Learning (TEL 2013) – P. 463-469.

[16] A.E. Yankovskaya, A.I. Gedike, R.V. Ametov, A.M. Bleikher IMSLOG-2002 software tool for supporting information technologies of test pattern recognition. Pattern Recognition and Image Analysis, 2003. Vol. 13. No. 4. – P. 650-657.

[17] A.E. Yankovskaya Transformation of features space in patterns space on the base of the logical-combinatorial methods and properties of some geometric figures. Proceedings of the International Conference Pattern Recognition and Image Analysis: New Information, Abstracts of the I All-Union Conference, Part II, pp. 178-181,Minsk, 1991.(in Russian).

[18] S.V. Kondratenko, A.E. Yankovskaya System of visualization TRIANG for decision-making justification with cognitive graphics usage. Proceedings of the Third Conference on Artificial Intelligence. Vol. I. - Tver, 1992. - p. 152-155 (in Russian).

[19] A.V. Yamshanov, N.M. Krivdyuk Specify of software implementation of cognitive graphic tools in intelligent and education systems. Proceedings of Fundamental Science Development XI, Russia, Tomsk, 22–25 April 2014, (ISBN 978-5-4387-0415-7) (in Russian).

[20] A.E. Yankovskaya An Automaton Model, Fuzzy Logic, and Means of Cognitive Graphics in the Solution of Forecast Problems. Pattern Recognition and Image Analysis. – 1998. – Vol. 8, No. 2. – pp. 154-156.

[21] A.E. Yankovskaya, Y.A. Shurigin, A.V. Yamshanov, N.M. Krivdyuk Determination of the student knowledge level on the base of a learning course which is presented by a semantic network. Procedings of Open Semantic Technologies for Intelligent Systems (OSTIS-2015). – Minsk : BSUIR, p. 331-339, 2015 (in Russian).

[22] Andreea Mateescua, Mihaela Chraif. The relationship between job satisfaction, occupational stress and coping mechanism in educational and technical organizations. Procedia - Social and Behavioral Sciences 187 (2015) 728 – 732.

[23] Cătălina Dumitrescu. Reducing the self-perceived stress level, heart rate and blood pressure by cognitive behavioral intervention plan in a multinational organization from Romania. Procedia - Social and Behavioral Sciences 187 (2015) 704 – 707.

[24] Cristiana Catalina Ciceia. Occupational stress and organizational commitment in Romanian public organizations. Procedia - Social and Behavioral Sciences 33 (2012) 1077 – 1081.

[25] Alan Patchinga, Rick Best. An investigation into psychological stress detection and management in organisations operating in project and construction management. Procedia - Social and Behavioral Sciences 119 ( 2014 ) 682 – 691.

[26] Che Noriah Othman, Roz Azinur Che Lamin, Nursyuhadah Othman. Occupational Stress Index of Malaysian University. Procedia - Social and Behavioral Sciences 153 (2014) 700 – 710.

[27] Ruxandra Foloútină, Loredana Adriana Tudorache. Stress management tools for preventing burnout phenomenon at teachers from special education. International Conference on Education and Educational Psychology (ICEEPSY 2012). Procedia - Social and Behavioral Sciences 69 ( 2012 ) 933 – 941.

[28] José Carlos Teixeira et al. Cognitive stimulation, maintenance and rehabilitation. HCIST 2013 - International Conference on Health and Social Care Information Systems and Technologies. Procedia Technology 9 ( 2013 ) 1335 – 1343.

[29] José Navarro, Robert A. Roe, María I. Artiles. Taking time seriously: Changing practices and perspectives in Work/Organizational Psychology. Journal of Work and Organizational Psychology 31 (2015) 135–145.

[30] Aharon Tziner, Edna Rabenu, Ruth Radomski, Alexander Belkin. Work stress and turnover intentions among hospital physicians: The mediating role of burnout and work satisfaction. ournal of Work and Organizational Psychology 31 (2015) 207–213.

[31] Nicole R. Fowlera et al. Cognitive testing in older primary care patients: A cluster-randomized trial. Alzheimer's & Dementia: Diagnosis, Assessment & Disease Monitoring 1 (2015) 349-357.

[32] A.E. Yankovskaya, S.V. Kitler, A.V. Silaeva Intelligent system of diagnostics and intervention of organizational stress : its development and testing. Open education 2012. – No 2 (91). – P. 61-69 (in Russian).

[33] N.A. Kornetov, A.E. Yankovskaya, S.V. Kitler, A.V. Silaeva, L.V. Shagalova bout development dynamic of representations about organizational stress and approaches to its evaluation. Fundamental Research №10, 2011 (in Russian).

[34] Zadeh Lotfi A. Fuzzy Logic, Neural Networks, and Soft Computing. Communications of the ACM. – Vol. 37. – N. 3. – P. 77-84..

[35]  Selye H. A Syndrome Produced by Diverse Nocuous Agents. Nature. – 1936. Vol. 138, – P. 32.
[36]  A.E. Yankovskaya, A.V. Yamshanov, N.M. Krivdyuk Cognitive graphic tools in intelligent learning-
      testing systems. Procedings of Open Semantic Technologies for Intelligent Systems (OSTIS-2014). –
      Minsk : BSUIR, p. 303 -308, 2014 (in Russian).

## AUTHORS

**Professor Anna Yankovskaya** obtained her DSc in Computer Science from the
Tomsk State University in Russia. She is currently a head of the Intelligent Systems
Laboratory and a professor of the Applied Mathematics Department at Tomsk State
University of Architecture and Building, a professor of the Computer Science
Department at Tomsk State University, a professor of Tomsk State University of
Control Systems and Radioelectronics and a professor Siberian State Medical
University. She is the author of more than 600 publications and 7 monographies. Her
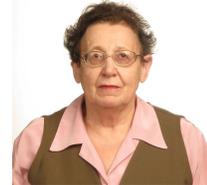
scientific interests include mathematical foundations for test pattern recognition and theory of digital
devices; artificial intelligence, intelligent systems, learning and testing systems, blended education and
learning; logical tests, mixed diagnostic tests, cognitive graphics; advanced technology in education.

**Artem Yamshanov** graduated from the Tomsk State University of Control Systems
and Radio Electronics in 2012. He is a postgraduate student at the Tomsk State
University of Control Systems and Radio Electronics. Research interests: intelligent
learning and testing systems, blended education, machine learning, artificial
intelligence,  intelligent systems and technologies data mining and pattern
recognition, cognitive tools and advanced technology in education.
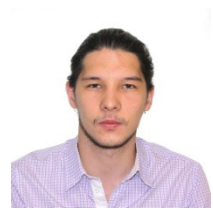